\documentclass[reprint,aps,amsmath,amssymb,amsfonts,showkeys]{revtex4-2}
\usepackage{graphicx}
\usepackage{dcolumn}
\usepackage{bm}
\usepackage{txfonts}
\usepackage{multirow}
\usepackage{color}

\begin{document}

\title{Third-order Electrical Conductivity of the Charge-ordered Organic Salt $\alpha$-(BEDT-TTF)$_2$I$_3$}

\author{Mayu~Ishii}
\author{Ryuji~Okazaki}
\email{okazaki@rs.tus.ac.jp}
\author{Masafumi~Tamura}
\affiliation{Department of Physics, Faculty of Science and Technology, Tokyo University of Science, Noda 278-8510, Japan}

\begin{abstract}

We performed 
third-order electrical conductivity measurements on the organic conductor 
$\alpha$-(BEDT-TTF)$_2$I$_3$
using an ac
bridge technique sensitive to nonlinear signals.
Third-order conductance $G_3$
is clearly observed even at low electric fields, and 
interestingly, $G_3$ is critically enhanced 
above the charge-order transition temperature $T_{\rm CO}=136$~K.
The observed frequency dependence of $G_3$ is incompatible with a percolation model,
in which a Joule heating in a random resistor network is relevant to the nonlinear conduction. 
We instead argue the nonlinearity of the relaxation time according to a phenomenological model on the mobility in materials with large dielectric constants,
and find that the third-order conductance $G_3$ corresponds to 
the third-order electric susceptibility $\chi_3$.
Since the nonlinear susceptibility is known as a probe for higher-order multipole ordering,
the present observation of the divergent behavior of $G_3$ above $T_{\rm CO}$ reveals
an underlying quadrupole
instability at the charge-order transition of the organic system.

\end{abstract}

\maketitle

%:intro
Electronic crystallization on a lattice,
known as charge ordering, is an 
emergent phenomenon driven by strong electron-electron interactions \cite{Seo2006,Hotta2012,Clay2019,Dressel2020}.
It is well materialized in 
a class of organic molecular conductors,
and 
a quasi-two-dimensional (quasi-2D) charge transfer salt $\alpha$-(ET)$_2$I$_3$
[ET being bis(ethylenedithio)-tetrathiafulvalene (BEDT-TTF)]
is known as a quite essential material to show charge ordering. 
$\alpha$-(ET)$_2$I$_3$
exhibits a metal-insulator transition at $T_{\rm CO}=136$~K \cite{Bender1984},
below which a stripe-type charge disproportionation (charge ordering) has been observed
by various experimental methods including 
nuclear magnetic resonance \cite{Takano2001,Takahashi2006}, 
optical measurements \cite{Meneghetti1986,Moldenhauer1993,Wojciechowski2003,Yue2010,Ivek2011,Yakushi2012}, 
and synchrotron radiation x-ray diffraction \cite{Kakiuchi2007}.
An electronic ferroelectricity without inversion symmetry is detected by
second harmonic generation technique \cite{Yamamoto2008} and 
the polarization switching is also confirmed \cite{Lunkenheimer2015}.
Interestingly, 
this charge-ordered state is highly sensitive to the external perturbations
to realize massless Dirac fermions in hydrostatic pressure and superconducting state under uniaxial pressure \cite{Tajima2002,Tajima2007,Kobayashi2007,Tajima2009}.

Among the various intriguing properties of $\alpha$-(ET)$_2$I$_3$,
the detailed response function for the electric field,
which may act as a conjugate field to the electronic crystal,
is of fundamental interest.
Indeed, charge ordering in $\alpha$-(ET)$_2$I$_3$ shows nontrivial behavior for applied electric field \cite{Dressel1995,Tamura2010,Itose2013,Ivek2012,Peterseim2016,Ito2013,Kodama2012,Uji2013,Iimori2007,Iimori2014,Ishikawa2014},
and moreover, a novel current-induced thermoelectric phenomenon has been also suggested recently \cite{Osada2021,Kiswandhi2021}.
Dressel \textit{et al.} first reported a nonlinear electrical conductivity above the threshold fields \cite{Dressel1995},
reminiscent of sliding motion of charge density wave (CDW).
Voltage oscillation and Shapiro steps have been subsequently observed above the threshold fields \cite{Tamura2010,Itose2013},
indicating a collective excitation of charge ordering similar to CDW \cite{Ivek2012}.
In addition, nonequilibrium hot-electron state is recently proposed by simultaneous optical and transport measurements at high fields \cite{Peterseim2016},
while
a scanning Raman microspectroscopy experiment has clearly shown that 
application of such a high electric field triggers 
a filamentary-like spatially-inhomogeneous state \cite{Ito2013},
which often becomes an extrinsic origin of
nonlinear conduction \cite{Okazaki2013}.
At low fields where such a spatial inhomogeneity is irrelevant,
a nonlinear conduction is also examined and 
the observed power-law current($I$)-voltage($V$) curves ($I\propto V^{\alpha}$) are explained by 
thermal excitation of electron-hole pairs from 2D logarithmic potential \cite{Kodama2012}.
In particular, 
the power-law exponent $\alpha$ in low-field range abruptly changes from $\alpha=1$ to 3 below $T_{\rm KT}\approx 35$~K,
suggesting 
an occurrence of a Kosterlitz-Thouless transition at $T_{\rm KT}$ \cite{Uji2013}.
In contrast, 
above
$T_{\rm KT}$,
an Ohmic $I$-$V$ curve ($\alpha=1$) is obtained at low fields even in the charge-ordered phase; 
a crucial issue how the nonlinear response emerges with the formation of the charge ordering at $T_{\rm CO}$ has not been discussed.

The aim of this paper is to clarify such an unexplored relation in $\alpha$-(ET)$_2$I$_3$ by means of the
third-order electrical conductivity measurements.
This harmonic method enables us to directly measure the nonlinear conductivity in low-field range
by detecting $3\omega$ component for the applied voltage with a fundamental frequency $\omega$,
in contrast to conventional pulsed technique in which the self-heating effect is unavoidable at high fields.
The third-order conductance $G_3$ has been observed in a whole temperature range across $T_{\rm CO}$ and
largely enhanced at $T_{\rm CO}$.
To explain this nonlinearity, we refer to the mobility of the polar materials recently investigated,
and find that the
third-order conductance $G_3$ approximately corresponds to 
the third-order electric susceptibility $\chi_3$, 
which expresses a quadrupole fluctuation.
Our present results thus imply a quadrupole nature in the charge-ordered organic salt $\alpha$-(ET)$_2$I$_3$.

\begin{figure}[t!]
\begin{center}
\includegraphics[width=1\linewidth]{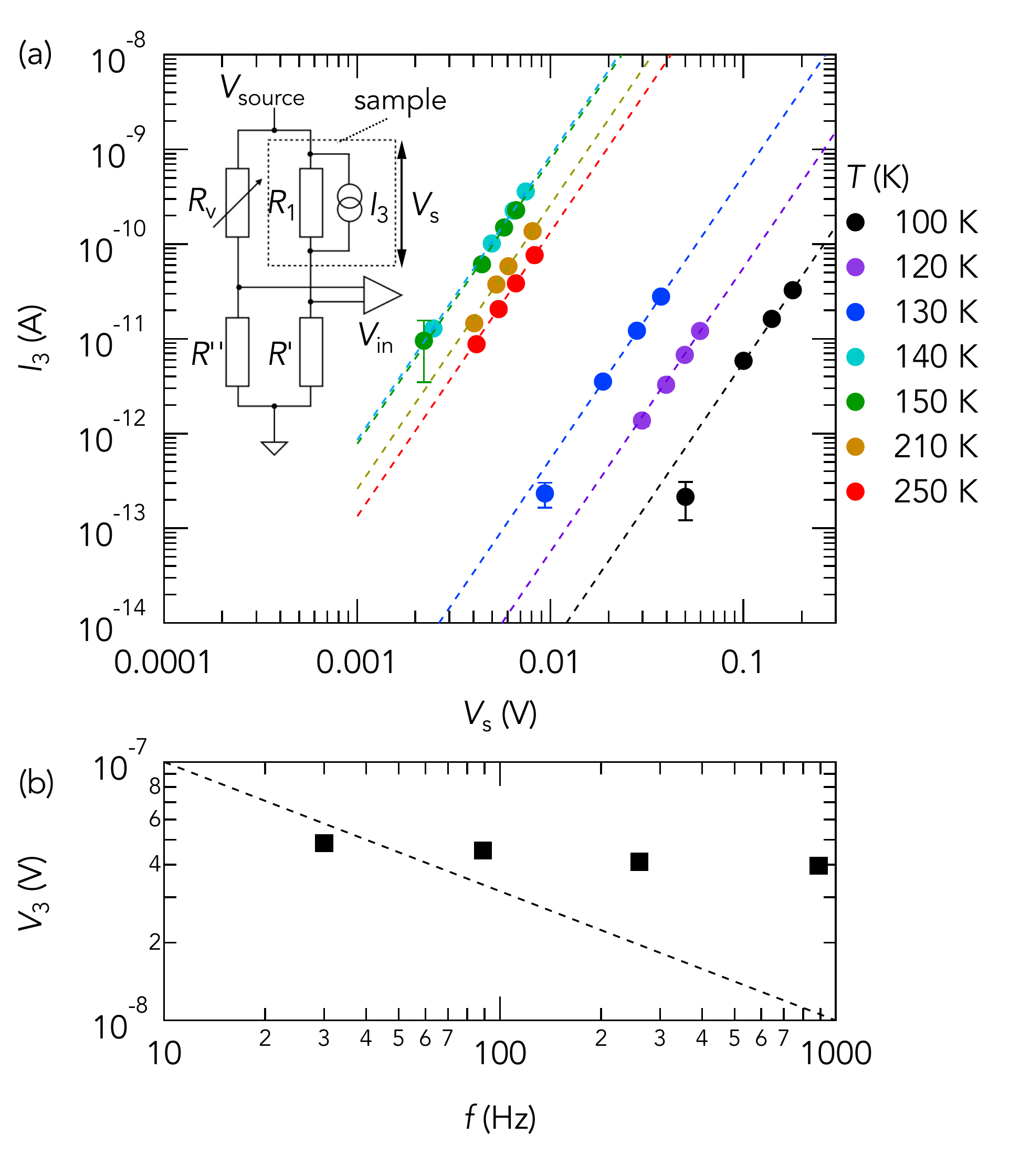}
\caption{
(Color online)
(a) In-phase components of third-harmonic current $I_3$ as a function of the applied voltage to the sample $V_{\rm s}$ 
measured at several temperatures.
The dashed lines represent the $V_{\rm s}^3$ dependence.
The inset depicts the bridge circuit to exclude the linear voltage into the lock-in amplifier.
(b) The frequency dependence of the third-harmonic voltage $V_3$ measured at $T=136$~K with the sample voltage of $V_{\rm s}\approx 6$~mV.
The dashed lines represent the $f^{-1/2}$ dependence, which is expected in the case that the self-heating effect is dominant.}
\end{center}
\end{figure}

Single crystals of $\alpha$-(ET)$_2$I$_3$ were 
prepared by an electrochemical method.
Two gold wires were attached by using a gold paste to reduce the contact resistance \cite{Tajima2000}
and the voltage was applied along 
the in-plane direction.
The rate of  temperature change 
is lower than 0.3~K/min
to prevent the damage to the sample.
As schematically shown in the inset of Fig. 1(a), 
third-order electrical conductivity measurements were performed by using a bridge circuit
to exclude the extrinsic third-order terms coming from nonlinearities of voltage source and 
electrical circuit inside the lock-in amplifier \cite{Gu1965,Dubson1989,Thibierge2008}.
When a sample exhibits nonlinear conductivity, 
electrical current $I$ flowing in the sample under the applied voltage $V_{\rm s}$, is expressed
by a Taylor series as 
\begin{align}
I \simeq I_1 + I_3 = G_1V_{\rm s} + G_3V_{\rm s}^3,
\end{align}
where $I_1$ ($G_1$) and $I_3$ ($G_3$) are linear and third-order current (conductance), 
respectively \cite{Thibierge2008}, and the higher-order terms are assumed to be negligible. 
Note that the inversion symmetry is broken below $T_{\rm CO}$ in $\alpha$-(ET)$_2$I$_3$ 
but we have not applied a magnetic field
in this experiment, 
so the second-order term $G_2V_{\rm s}^2$ is not considered here \cite{note1}.
Thus, the sample can be regarded as a parallel circuit 
consisting of
a linear resistance $R_1=G_1^{-1}$ and a nonlinear current source $I_3$.

In the bridge circuit, the input voltage to the lock-in amplifier $V_{\rm in}$ is then given as 
\begin{align}
V_{\rm in} = \frac{R_1R''-R_{\rm v}R'}{(R_1+R')(R_{\rm v}+R'')}V_{\rm source}
-\frac{I_3}{1/R_1+1/R'},
\label{bridge}
\end{align}
where $R_{\rm v}$, $R'$, and $R''$ are resistances 
constituting
the bridge circuit
and $V_{\rm source}$ is the source voltage \cite{Thibierge2008}. 
The voltage signal measured by a lock-in amplifier $V_{\rm m}$ must contain 
small but finite nonlinear components as 
$
V_{\rm m} \simeq V_{\rm in}\left\{1 + \beta(V_{\rm in})\right\},
$
where non-dimensional $\beta(V_{\rm in})$ is a series of $V_{\rm in}$ to express the nonlinearity in the lock-in amplifier,
order of which is typically $10^{-4}$ \cite{Thibierge2008}.
Here, in the balanced condition of $R_1R''=R_{\rm v}R'$, 
the first term in the right-hand side of Eq. (\ref{bridge}) becomes zero, and then 
the nonlinearity of the voltage source can be excluded.
In addition, in the balanced condition, one can also eliminate the nonlinearity of the lock-in amplifier
since the third-order term, $-I_3/(1/R_1+1/R')$, becomes the leading term.
Moreover, in the measured frequency range,
the frequency dependence of the linear conductivity is negligible, i.e. $G_1(\omega)=G_1(3\omega)$ \cite{Lunkenheimer2015}, so that
the balanced condition is kept both in $\omega$ and $3\omega$.
Note that
this harmonic technique is similar to the $3\omega$ method for thermal conductivity measurements \cite{Cahill1987,Moon1996}
and the ac method for the specific heat measurements \cite{Sullivan1968,Jung1992}, in which the harmonic signals from the heater and 
the thermometer are respectively measured, while
the present method detects the third-harmonic voltage of the sample directly.

In this experiments, we applied the ac voltage 
$V_{\rm source}(t) = V_0\sin \omega t$ 
to the circuit
with the frequency of $f = \omega/2\pi = 89$~Hz, and measured the fundamental 
voltage $V_1$ and the third-harmonic voltage $V_3$ simultaneously 
as 
$V_{\rm m} = V_1\sin\omega t+V_3\sin3\omega t$
by using two lock-in amplifiers to detect $1\omega$ and $3\omega$ signals.
We then balance the circuit by adjusting  variable resistance $R_v$ 
to achieve $V_1=0$, and obtain $I_3$ by using measured $V_3$ as
\begin{align}
I_3  = -(1/R_1+1/R')V_3= G_3V_{\rm s}^3,
\label{thirdcurrent}
\end{align}
where $V_{\rm s} \simeq V_0R_1/(R_1+R')$ is the applied voltage to the sample \cite{Thibierge2008}.
It is worth to note that the measured voltage ratio $V_3/V_1$ is 
as small as the order of $10^{-6}$, 
which demonstrates the necessity of the bridge technique to exclude extrinsic nonlinearities as mentioned above.
We also utilized a four-terminal double-bridge method \cite{Dubson1989} for the low-resistance range and the results were 
similar to that of the two-terminal bridge method.

\begin{figure}[t!]
\begin{center}
\includegraphics[width=1\linewidth]{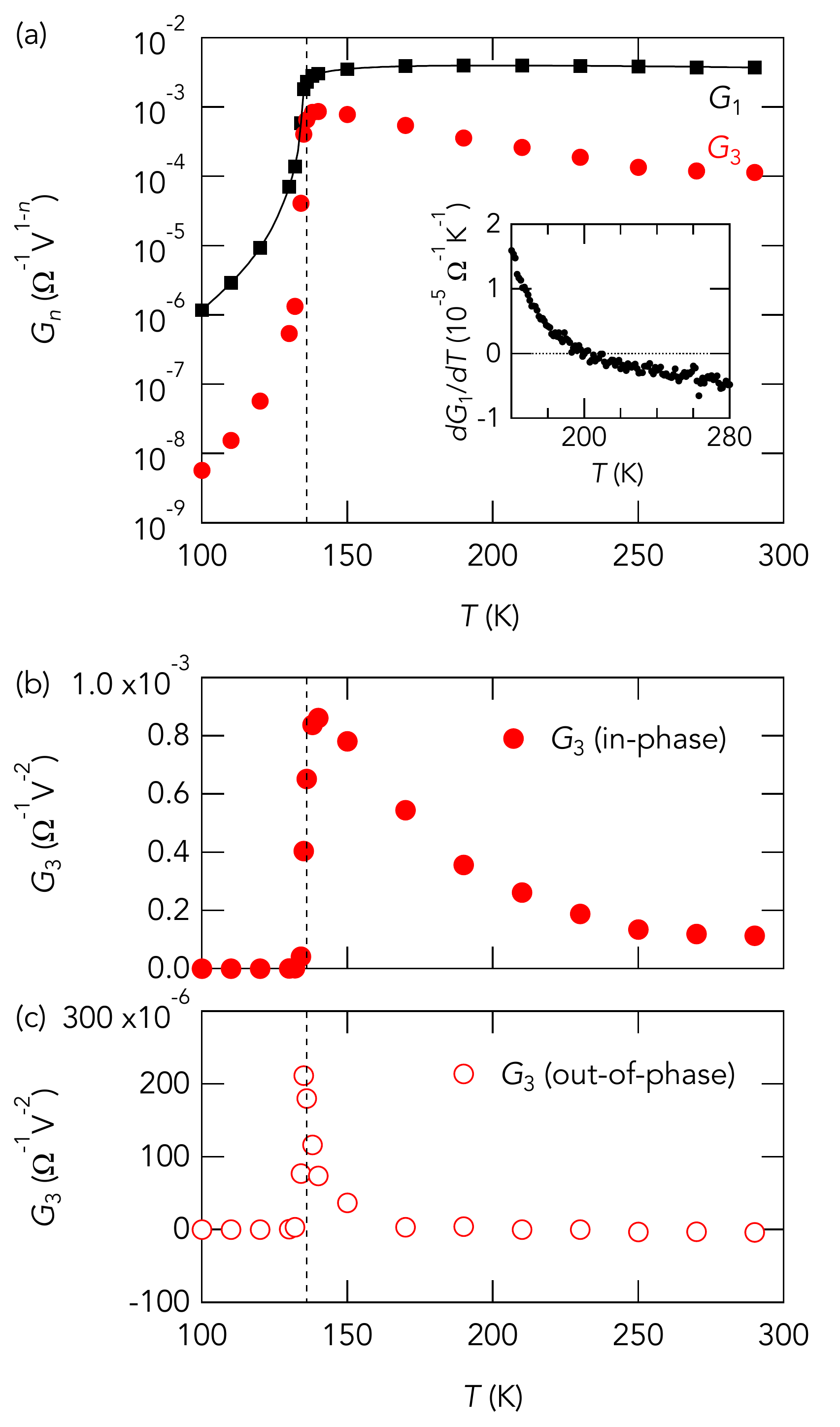}
\caption{
(Color online)
(a) Temperature variations of the linear conductance $G_1$ (squares) and the in-phase component of the third-harmonic conductance $G_3$ (circles).
Solid curve is the linear conductance measured with a constant current of $I=$ 10~nA.
The inset shows temperature dependence of $dG_1/dT$.
(b,c) In-phase and out-of-phase components of the third-harmonic conductance $G_3$ as a function of temperature.
Dashed line represents the charge order transition temperature $T_{\rm CO} = $136~K.
}
\end{center}
\end{figure}

Figure 1(a) depicts the in-phase component of the third-harmonic current $I_3$ as a function of $V_{\rm s}$
obtained at several temperatures. 
Hereafter all the current and voltage data are expressed by rms.
Although the signals are comparable to the noise level at the low-current range below $10^{-12}$ A,
which is close to the noise level in the earlier report \cite{Uji2013},
$I_3$ is certainly proportional to $V_{\rm s}^3$, and the slope gives the 
third-order conductance $G_3$ according to Eq. (\ref{thirdcurrent}).
We emphasize that 
the electric field corresponding the present voltage is lower than 10 V/cm at $T=100$~K, 
which is fairly low compared with the earlier reports \cite{Tamura2010,Itose2013,Peterseim2016}.

Figure 2(a) compares temperature dependence of the fundamental and 
third-order conductances $G_1$ and $G_3$, both of which are the in-phase components.
Note that, although we utilize the two-terminal method owing to the usage of bridge circuit,
the obtained temperature variation of $\sigma_1=G_1\times (L/S)$,
where $L\approx 4\times 10^{-2}$ cm and $S\approx 1\times 10^{-5}$ cm$^2$ are the length between the terminals and the cross-sectional area respectively,
is close to that measured by the
four-terminal method, implying that the effect of the contact resistance formed by the gold paint is negligible.
Figures 2(b) and 2(c) represent the temperature variations of the in-phase and out-of-phase components of 
third-order conductance $G_3$, respectively.
Interestingly, $G_3$ is observed even above $T_{\rm CO} = 136$~K, and
compared with temperature dependence of $G_1$, 
$G_3(T)$ is notably enhanced near $T_{\rm CO}$, as is discussed in the following section.

\begin{table}[b]
  \begin{center}
    \caption{A simple sort of origins for the nonlinear conduction in low field range.}
    \begin{tabular}{cccc} 
    \hline
      \multirow{4}{*}{\:$\sigma(E)$\:} & \multirow{2}{*}{\:$\sigma[T(E)]$\:} & \:$\sigma[T_{\rm p}(E)]$\: & \: self-heating effect \cite{Dubson1989} \: \\ \cline{3-4}    
      &  & $\sigma[T_{\rm e}(E)]$ & hot electron \cite{Ryder1953} \\ \cline{2-4}    
      & \multirow{2}{*}{$\sigma(E)$} & $en(E)\mu$ & \: field-dependent energy gap \cite{Kodama2012,Uji2013} \: \\ \cline{3-4}   
       &  & $en\mu(E)$ & field-dependent mobility \\ \hline
    \end{tabular}
  \end{center}
\end{table}

Let us consider the origin of the observed third-order nonlinearity.
Although there are various causes of the nonlinear conduction, 
as categorized in Table I, 
one may first sort it whether the conductivity $\sigma$ varies when the temperature changes.
Note that the applied electric field to the sample is lower
than the threshold field \cite{Tamura2010,Itose2013}, 
above which
charge ordering shows a collective sliding mode as is similar to that of CDW \cite{Gruner}.
It should also be emphasized that there are few studies on the third-harmonic conductivity \cite{Mosh2009,Rommel2013} and 
the interpretation remains unclear.
In the simplest case, the phonon (lattice) temperature $T_{\rm p}$ 
increases
to
$T_{\rm p}+\Delta T_{\rm p}$ by the applied power of $G_1V^2$,
and then the conductivity seems to be varied,
which is well known as the extrinsic self-heating effect.
In this case, the conductance is expressed as 
$
G(V) \simeq G_1 + (\partial G_1/\partial T_{\rm p})\Delta T_{\rm p}
$
and the temperature increase is given as $\Delta T_{\rm p} = R_{\rm th}G_1V^2$, where
$R_{\rm th}$ is the thermal resistance between the sample and the thermal bath,
leading to
the third-order conductance of $G_3 = R_{\rm th}G_1(\partial G_1/\partial T_{\rm p})$ \cite{Dubson1989}.
This heating effect is unlikely in the present case, because
the sign of observed $G_3$ is positive while $\partial G_1/\partial T_{\rm p}$ changes its sign
at around 200 K as seen in the inset of Fig. 2(a).
Note that 
this nonlinearity due to the heating effect is of importance
to detect higher-order moment of the current distribution in the percolation network
consisting of the metallic and insulating domains in the sample. 
In this case,
a peculiar frequency dependence of $V_3\propto f^{-1/2}$ is expected \cite{Dubson1989}.
On the other hand, 
the observed $V_3$ shows almost no frequency dependence as shown in Fig. 1(b),
again indicating that the heating effect is unlikely in the present case.

We then discuss the hot-electron model. In this scenario, an electron temperature $T_{\rm e}$,
which is larger than $T_{\rm p}$, is enhanced by applying electric field to vary the conductivity.
This model is suggested to explain the high-field nonlinearity in this material \cite{Peterseim2016}.
In this case, the electron and phonon systems are connected with the thermal resistance $R_{\rm ep}$
and the characteristic time scale $\tau_{\rm e}$ is given as $\tau_{\rm e}=R_{\rm ep}C_{\rm e}$, where
$C_{\rm e}$ is the electronic specific heat \cite{Ryder1953}.
Now the order of $\tau_{\rm e}$ is generally as small as $10^{-9}$~s, 
and thus previous hot-electron phenomena have been investigated by using a pulsed technique 
with a short pulse width of $10^{-9}$~s \cite{Ryder1953}.
In contrast, the period of the applied voltage in the present measurement of $f^{-1}\sim 10^{-2}$~s
is much longer than $\tau_{\rm e}$.
Therefore, 
the electron and lattice systems are in equilibrium ($T_{\rm e} \simeq T_{\rm p}$) in the present time scale,
and thus hot-electron scenario is excluded due to the same reason to the self-heating case as discussed above.

Next we argue a non-thermal origin for nonlinear conduction, 
which is not induced through the temperature change in field.
If we adapt a simple Drude formula of $\sigma=en\mu$ for the measured temperature range, 
where $n$ and $\mu$ are the carrier density and the mobility, respectively,
one may consider 
the electric-field dependence of $n$ and/or $\mu$ for the nonlinearity of $\sigma$.
According to Refs. \onlinecite{Kodama2012,Uji2013},
the charge gap $\Delta$ is effectively reduced by the field $E$ as
$\Delta (E) =U_0\ln(\lambda/a)-e\lambda E$,
where $\lambda$ is a cutoff length of the potential and $a$ is the minimum length scale taken as the size of the ET molecule,
resulting in increase of the activation-type conductivity 
$\sigma(E) = \sigma_0\exp[-\Delta(E)/2k_{\rm B}T]$ due to the 
field-induced increase of the carrier density $n(E)$.
However, this scenario is based on the 2D logarithmic potential developed well below $T_{\rm CO}$,
in contrast to the present results that the nonlinearity appears critically above $T_{\rm CO}$.

\begin{figure}[t!]
\begin{center}
\includegraphics[width=1\linewidth]{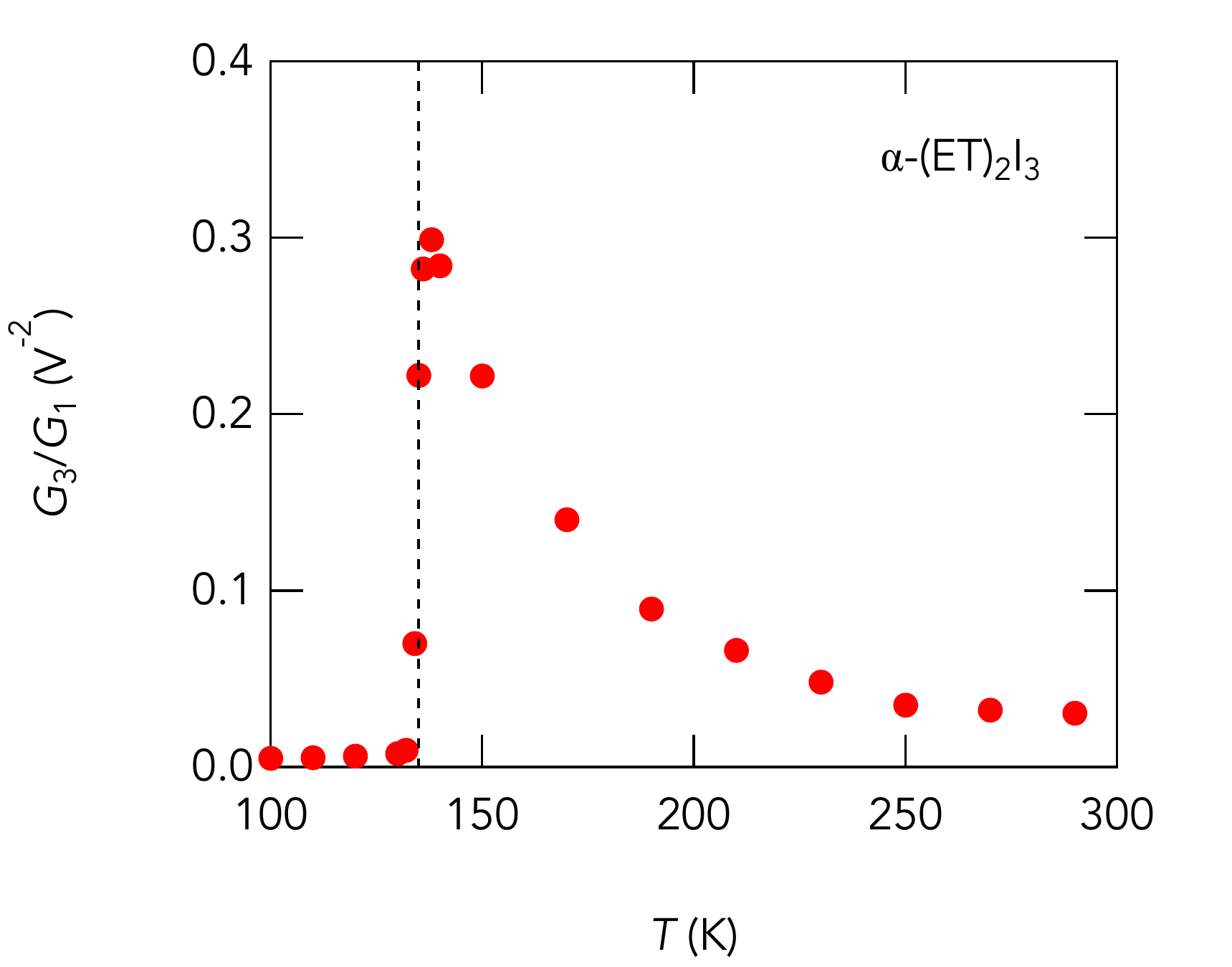}
\caption{
(Color online)
Temperature dependence of the ratio of the in-phase components of $G_3$ to $G_1$.
Dashed line represents the charge order transition temperature $T_{\rm CO} = $136~K.
}
\end{center}
\end{figure}

We then consider possible field-dependent mobility $\mu(E)$.
Recently, it has been suggested that 
the mobility in the large-permittivity compound is expressed as
$\mu\propto\varepsilon^{1/2}$ ($\varepsilon$ is the permittivity) \cite{Behnia2015,Wang2019}.
Here, mean free path $l$ in large-permittivity materials
may be proportional to the Thomas-Fermi screening length $r_{\rm TR}$,
since it gives a characteristic length scale for the potential modulation which acts as the scattering centers \cite{Behnia2015}.
In such a case, since $r_{\rm TR}\propto \sqrt{a_{\rm B}^*}$, where $a_{\rm B}^*$ is the effective Bohr radius given as 
$a_{\rm B}^*=(\varepsilon/\varepsilon_0)(m/m^*)a_{\rm B}$ ($\varepsilon_0$ being the vacuum permittivity, $m$ the electron mass,
$m^*$ the effective mass, $a_{\rm B}$ the Bohr radius), 
the mobility $\mu$ is given as $\mu\propto l\propto \varepsilon^{1/2}$.
Here, the permittivity is given as $\varepsilon=\varepsilon_0(1+\chi)$ and the 
electric susceptibility $\chi$
is generally expanded as $\chi \simeq \chi_1+\chi_3E^2$, where
$\chi_1$ and $\chi_3$ are linear and third-order electric susceptibilities, respectively.
Up to the $E^2$ term, 
we then obtain a field-dependent mobility as
\begin{align}
\mu(E) \propto (1+\chi)^{1/2}
\simeq(1+\chi_1)^{1/2}\left(1+\frac{1}{2}\frac{\chi_3}{1+\chi_1}E^2\right).
\label{perm1}
\end{align}
Therefore, by comparing it with $\sigma=en\mu \simeq \sigma_1[1+(\sigma_3/\sigma_1)E^2]$, the conductance ratio is given as
\begin{align}
\frac{G_3}{G_1}\propto\frac{\sigma_3}{\sigma_1}\propto
\frac{1}{2}\frac{\chi_3}{1+\chi_1}.
\label{ratio2}
\end{align}
In Fig. 3, we plot temperature dependence of $G_3/G_1$. 
Interestingly, it
exhibits a peak structure near $T_{\rm CO}$, indicating an 
enhancement of the third-order electric susceptibility $\chi_3$ at the charge-order transition.
Note that the linear susceptibility $\chi_1$ in the denominator is difficult to contribute to the enhancement of $G_3/G_1$,
because $\chi_1$ is usually enhanced near the charge-order transition as seen in the one-dimensional organic salts \cite{Nad2006}
while
it is very difficult to measure $\chi_1$ above $T_{\rm CO}$ in $\alpha$-(ET)$_2$I$_3$ owing to the high conductivity \cite{Lunkenheimerreview}.
We also mention that
the observed enhancement of the out-of-phase component of $G_3$ at $T_{\rm CO}$ [Fig. 2(c)] 
may support this picture
because 
the imaginary part of the complex conductivity $\tilde{\sigma}=\sigma+i\omega\chi\varepsilon_0$
generally expresses the electric susceptibility.

In general,
third-order electric susceptibility thermodynamically represents the fluctuation of the 
quadrupole moments \cite{Mart1996}.
In the magnetic analogy, the higher-order magnetic susceptibility is known as the probe for multipole ordering \cite{Morin1981,Ramirez1992}.
Therefore,
the present results imply a quadrupole instability hidden in the charge-ordered salt $\alpha$-(ET)$_2$I$_3$,
which may be natural in a sense that it is also an even-ordered multipole expansion.
It should be noted that
a signature of charge order is observed with an inhomogeneity even in the high-temperature metallic phase \cite{Yue2010,Katano2015}.
Also note that such a quadrupole nature is mentioned in $\kappa$-type ET salt \cite{Rommel2013} and manganese oxide \cite{Mosh2009}, 
although these nonlinear experiments were performed with a 
standard lock-in technique without bridge circuit in contrast to the present study.
Now the quadrupole nature in $\alpha$-(ET)$_2$I$_3$ may be attributed to two possible origins: 
Firstly, the charge-density distribution should be asymmetric in the ET molecule, leading to the quadrupole moment.
This is supported by the fact that the framework of ET molecule is sensitive to the amount of charge \cite{Umland1988}.
Note that the precise determination of the molecular structure using the synchrotron radiation is recently developed \cite{Kitou2017,Kitou2020}.
Secondly, the charge distribution inside the unit cell becomes more asymmetric by the formation of charge ordering at $T_{\rm CO}$.
This is also deduced from the temperature dependence of the in-plane anisotropy of the resistivity \cite{Ivek2010,Ivek2017}.
At present, it is difficult 
to clarify the most appropriate interpretation;
both origins seem to contribute to the quadrupole instability.

To summarize, we have measured the third-order conductivity of $\alpha$-(ET)$_2$I$_3$ over a broad temperature range across the 
charge ordering transition temperature $T_{\rm CO}=136$~K.
We find the enhancement of the third-order conductance at $T_{\rm CO}$, which may attribute to 
the possible quadrupole instability hidden in the charge-ordered organic salt.

\begin{acknowledgments}
We thank M. D. Anoop, N. Kikugawa, Y. Maeno, I. Terasaki, K. Ueda, and S. Yonezawa for discussion.
This work was supported by JSPS KAKENHI Grants No. 17H06136.
\end{acknowledgments}

%:ref
%%%%%%%%%%%%%%%%%%%%%%%%%%%%%%%%%%%%%%%%%%%%%%%%%%%%%%%%%%%%%%%%%%%%%%%%%%%%%%%%%
%\input{reference}
%\section{reference}


\begin{thebibliography}{99}

\bibitem{Seo2006}
H. Seo, J. Merino, H. Yoshioka, and M. Ogata,
J. Phys. Soc. Jpn. {\bf 75}, 051009 (2006).

\bibitem{Hotta2012}
C. Hotta, 
Crystals {\bf 2}, 1155 (2012).

\bibitem{Clay2019}
R. T. Clay and S. Mazumdar,
Phys. Rep. {\bf 788}, 1 (2019).

\bibitem{Dressel2020}
M. Dressel and S. Tomi\'{c}, Adv. Phys. {\bf 69}, 1 (2020).

\bibitem{Bender1984}
K. Bender, I. Hennig, D. Schweitzer, K. Dietz, H. Endres, and H. J. Keller, 
Mol. Cryst. Liq. Cryst. {\bf 108}, 359 (1984).

\bibitem{Takano2001}
Y. Takano, K. Hiraki, H. Yamamoto, T. Nakamura, and T. Takahashi, 
J. Phys. Chem. Solids {\bf 62}, 393 (2001).

\bibitem{Takahashi2006}
T. Takahashi, Y. Nogami, and K. Yakushi, 
J. Phys. Soc. Jpn. {\bf 75}, 051008 (2006).

\bibitem{Meneghetti1986}
M. Meneghetti, R. Bozio, and C. Pecile,
Journal de Physique {\bf 47}, 1377 (1986).

\bibitem{Moldenhauer1993}
J. Moldenhauer, C. Horn, K. Pokhodnia, D. Schweitzer, I. Heinen, and H. Keller, 
Synth. Met. {\bf 60}, 31 (1993).

\bibitem{Wojciechowski2003}
R. Wojciechowski, K. Yamamoto, K. Yakushi, M. Inokuchi, and A. Kawamoto, 
Phys. Rev. B {\bf 67}, 224105 (2003).

\bibitem{Yue2010}
Y. Yue, K. Yamamoto, M. Uruichi, C. Nakano, K. Yakushi, S. Yamada, T. Hiejima, and A. Kawamoto, 
Phys. Rev. B {\bf 82}, 075134 (2010).

\bibitem{Ivek2011}
T. Ivek, B. Korin-Hamzi\'{c}, O. Milat, S. Tomi\'{c}, C. Clauss, N. Drichko, D. Schweitzer, and M. Dressel, 
Phys. Rev. B {\bf 83}, 165128 (2011).

\bibitem{Yakushi2012}
K. Yakushi, 
Crystals {\bf 2}, 1291 (2012).

\bibitem{Kakiuchi2007}
T. Kakiuchi, Y. Wakabayashi, H. Sawa, T. Takahashi, and T. Nakamura, 
J. Phys. Soc. Jpn. {\bf 76}, 113702 (2007).

\bibitem{Yamamoto2008}
K. Yamamoto, S. Iwai, S. Boyko, A. Kashiwazaki, F. Hiramatsu, C. Okabe, N. Nishi, and K. Yakushi, 
J. Phys. Soc. Jpn. {\bf 77}, 074709 (2008).

\bibitem{Lunkenheimer2015}
P. Lunkenheimer, B. Hartmann, M. Lang, J. M\"uller, D. Schweitzer, S. Krohns, and A. Loidl,
Phys. Rev. B {\bf 91}, 245132 (2015).

\bibitem{Tajima2002}
N. Tajima, A. Ebina-Tajima, M. Tamura, Y. Nishio, and K. Kajita, 
J. Phys. Soc. Jpn. {\bf 71}, 1832 (2002).

\bibitem{Tajima2007}
N. Tajima, S. Sugawara, M. Tamura, R. Kato, Y. Nishio, and K. Kajita, 
Eur. Phys. Lett. {\bf 80}, 47002 (2007).

\bibitem{Kobayashi2007}
A. Kobayashi, S. Katayama, Y. Suzumura, and H. Fukuyama,
J. Phys. Soc. Jpn. {\bf 76}, 034711 (2007).

\bibitem{Tajima2009}
N. Tajima, S. Sugawara, R. Kato, Y. Nishio, and K. Kajita, 
Phys. Rev. Lett. {\bf 102}, 176403 (2009).



\bibitem{Dressel1995}
M. Dressel, G. Gr\"{u}ner, J. Pouget, A. Breining, and D. Schweitzer, 
Synth. Met. {\bf 70}, 929 (1995).



\bibitem{Tamura2010}
K. Tamura, T. Ozawa, Y. Bando, T. Kawamoto, and T. Mori,
J. Appl. Phys. {\bf 107}, 103716 (2010).

\bibitem{Itose2013}
F. Itose, T. Kawamoto, and T. Mori,
J. Appl. Phys. {\bf 113}, 213702 (2013).

\bibitem{Ivek2012}
T. Ivek, I. Kova\ifmmode \check{c}\else \v{c}\fi{}evi\'{c}, M. Pinteri\'{c}, B. Korin-Hamzi\'{c}, S. Tomi\'{c}, T. Knoblauch, D. Schweitzer, and M. Dressel,
Phys. Rev. B {\bf 86}, 245125 (2012).

\bibitem{Peterseim2016}
T. Peterseim, T. Ivek, D. Schweitzer, and M. Dressel,
Phys. Rev. B {\bf 93}, 245133 (2016).

\bibitem{Ito2013}
A. Ito, Y. Nakamura, A. Nakamura, and H. Kishida,
Phys. Rev. Lett. {\bf 111}, 197801 (2013).

\bibitem{Kodama2012}
K. Kodama, M. Kimata, Y. Takahide, N. Kurita, A. Harada, H. Satsukawa, T. Terashima, S. Uji, K. Yamamoto, and K. Yakush,
J. Phys. Soc. Jpn. {\bf 81}, 044703 (2012).

\bibitem{Uji2013}
S. Uji, K. Kodama, K. Sugii, Y. Takahide, T. Terashima, N. Kurita, S. Tsuchiya, M. Kohno, M. Kimata, K. Yamamoto, and K. Yakushi,
Phys. Rev. Lett. {\bf 110}, 196602 (2013).

\bibitem{Iimori2007}
T. Iimori, T. Naito, and N. Ohta,
J. Am. Chem. Soc. {\bf 129}, 3486 (2007).

\bibitem{Iimori2014}
T. Iimori and N. Ohta, 
J. Phys. Chem. C {\bf 118}, 7251 (2014).

\bibitem{Ishikawa2014}
T. Ishikawa, Y. Sagae, Y. Naitoh, Y. Kawakami, H. Itoh, K. Yamamoto, K. Yakushi, H. Kishida, T. Sasaki, S. Ishihara, Y. Tanaka, K. Yonemitsu, and S. Iwai,
Nat. Commun. {\bf 5}, 5528 (2014).

\bibitem{Osada2021}
T. Osada and A. Kiswandhi,
J. Phys. Soc. Jpn. {\bf 90}, 053704 (2021).

\bibitem{Kiswandhi2021}
A. Kiswandhi and T. Osada, 
J. Phys.: Condens. Matter {\bf 34}, 105602 (2022).

\bibitem{Okazaki2013}
R. Okazaki, Y. Nishina, Y. Yasui, F. Nakamura, T. Suzuki, and I. Terasaki,
J. Phys. Soc. Jpn. {\bf 82}, 103702 (2013).

\bibitem{Tajima2000}
N. Tajima, M. Tamura, Y. Nishio, K. Kajita, and Y. Iye,
J. Phys. Soc. Jpn. {\bf 69}, 543 (2000).


\bibitem{Gu1965}
T. Guldbrandsen, N. I. Meyer, and J. Schjaer-Jakobsen,
Rev. Sci. Instrum. {\bf 36}, 743 (1965).

\bibitem{Dubson1989}
M. A. Dubson, Y. C. Hui, M. B. Weissman, and J. C. Garland, 
Phys. Rev. B {\bf 39}, 6807 (1989).

\bibitem{Thibierge2008}
C. Thibierge, D. L'H\^ote, F. Ladieu, and R. Tourbot,
Rev. Sci. Instrum. {\bf 79}, 103905 (2008).

\bibitem{note1}
Y. Tokura and N. Nagaosa,
Nat. Commun. {\bf 9}, 3740 (2018).

\bibitem{Cahill1987}
D. G. Cahill and R. O. Pohl,
Phys. Rev. B {\bf 35}, 4067 (1987).

\bibitem{Moon1996}
I. K. Moon, Y. H. Jeong, and S. I. Kwun,
Rev. Sci. Instrum. {\bf 67}, 29 (1996).

\bibitem{Sullivan1968}
P. F. Sullivan and G. Seidel,
Phys. Rev. {\bf 173}, 679 (1968).

\bibitem{Jung1992}
D. H. Jung, T. W. Kwon, D. J. Bae, I. K. Moon, and Y. H. Jeong,
Meas. Sci. Technol. {\bf 3}, 475 (1992).

\bibitem{Gruner}
G. Gr\"uner,
Rev. Mod. Phys. {\bf 60}, 1129 (1988);
Rev. Mod. Phys. {\bf 66}, 1 (1994).

\bibitem{Rommel2013}
R. Rommel, B. Hartmann, J. Brandenburg, J. A. Schlueter, and J. M\"uller,
Phys. Status Solidi B {\bf 250}, 568 (2013).

\bibitem{Mosh2009}
V. Moshnyaga, K. Gehrke, O. I. Lebedev, L. Sudheendra, A. Belenchuk, S. Raabe, O. Shapoval, J. Verbeeck, G. Van Tendeloo, and K. Samwer, 
Phys. Rev. B {\bf 79}, 134413 (2009).

\bibitem{Ryder1953}
E. J. Ryder,
Phys. Rev. {\bf 90}, 766 (1953).



\bibitem{Behnia2015}
K. Behnia, 
J. Phys.: Condens. Matter {\bf 27}, 375501 (2015).

\bibitem{Wang2019}
J. Wang, L. Yang, C. W. Rischau, Z. Xu, Z. Ren, T. Lorenz, J. Hemberger, X. Lin, and K. Behnia,
npj Quant. Mat. {\bf 4}, 61 (2019).

\bibitem{Nad2006}
F. Nad and P. Monceau,
J. Phys. Soc. Jpn. {\bf 75}, 051005 (2006).

\bibitem{Lunkenheimerreview}
P. Lunkenheimer and A. Loidl,
J. Phys.: Condens. Mat. {\bf 27}, 373001 (2015).

\bibitem{Mart1996}
R. Marto\ifmmode \check{n}\else \v{n}\fi{}\'ak and E. Tosatti,
Phys. Rev. B {\bf 54}, 15714 (1996).

\bibitem{Morin1981}
P. Morin and D. Schmitt,
Phys. Rev. B {\bf 23}, 5936 (1981).


\bibitem{Ramirez1992}
A. P. Ramirez, P. Coleman, P. Chandra, E. Br\"uck, A. A. Menovsky, Z. Fisk, and E. Bucher,
Phys. Rev. Lett. {\bf 68}, 2680 (1992).


\bibitem{Katano2015}
K. Katono, T. Taniguchi, K. Ichimura, Y. Kawashima, S. Tanda, and K. Yamamoto,
Phys. Rev. B {\bf 91}, 125110 (2015).


\bibitem{Umland1988}
T. C. Umland, S. Allie, T. Kuhlmann, and P. Coppens,
J. Chem. Phys. {\bf 92}, 6456 (1988).

\bibitem{Kitou2017}
S. Kitou, T. Fujii, T. Kawamoto, N. Katayama, S. Maki, E. Nishibori, K. Sugimoto, M. Takata, T. Nakamura, and H. Sawa,
Phys. Rev. Lett. {\bf 119}, 065701 (2017).

\bibitem{Kitou2020}
S. Kitou, Y. Hosogi, R. Kitaura, T. Naito, T. Nakamura, and H. Sawa,
Crystals {\bf 10}, 998 (2020).

\bibitem{Ivek2010}
T. Ivek, B. Korin-Hamzi\'c, O. Milat, and S. Tomi\'c, C. Clauss, N. Drichko, D. Schweitzer, and M. Dressel,
Phys. Rev. Lett. {\bf 104}, 206406 (2010).

\bibitem{Ivek2017}
T. Ivek, M. \ifmmode \check{C}\else \v{C}\fi{}ulo, M. Kuve\ifmmode \check{z}\else \v{z}\fi{}di\ifmmode \acute{c}\else \'{c}\fi{}, E. Tuti\ifmmode \check{s}\else \v{s}\fi{}, M. Basleti\ifmmode \acute{c}\else \'{c}\fi{}, B. Mihaljevi\ifmmode \acute{c}\else \'{c}\fi{}, E. Tafra, S. Tomi\ifmmode \acute{c}\else \'{c}\fi{}, A. L\"ohle, M. Dressel, D. Schweitzer, and B. Korin-Hamzi\ifmmode \acute{c}\else \'{c}\fi{}, 
Phys. Rev. B {\bf 96}, 075141 (2017).



\end{thebibliography}
\end{document}